\documentclass[fleqn,10pt]{wlscirep}
\usepackage{sistyle}

\title{Entropy spikes as a signature of Lifshitz transitions in the Dirac materials}

\author[1,+]{V.Yu.~Tsaran}
\author[2,3,+]{A.V. Kavokin}
\author[4,+,*]{S.G.~Sharapov}
\author[2,+]{A.A.~Varlamov}
\author[4,+]{V.P.~Gusynin}
\affil[1]{Institut f{\"u}r Kernphysik, Johannes Gutenberg Universit{\"a}t, D-55128 Mainz, Germany}
\affil[2]{CNR-SPIN, University ``Tor Vergata'', Viale del Politecnico 1, I-00133 Rome, Italy}
\affil[3]{Physics and Astronomy School, University of Southampton, Highfield, Southampton, SO171BJ, UK}
\affil[4]{Bogolyubov Institute for Theoretical Physics, National Academy of Science of Ukraine, 14-b
        Metrolohichna Street, Kiev 03680, Ukraine}

\affil[*]{sharapov@bitp.kiev.ua}

\affil[+]{these authors contributed equally to this work}

\begin{abstract}
We demonstrate theoretically that the characteristic feature of a 2D system undergoing $N$ consequent Lifshitz topological
transitions is the  occurrence of spikes of entropy per particle $s$ of
a magnitude $\pm  \ln 2/(J-1/2)$ with $2 \leq J \leq N$ at low temperatures.
We derive a general expression for $s$  as a function of chemical potential, temperature and  gap magnitude
for the gapped Dirac materials.
Inside the smallest gap, the dependence of $s$ on the chemical potential
exhibits  a dip-and-peak structure in the temperature vicinity of the Dirac point.
The spikes of the entropy per particles can be considered as a signature of the Dirac materials.
These distinctive characteristics of gapped Dirac materials can be detected
in transport experiments where the temperature is modulated in gated structures.
\end{abstract}
\begin{document}

\flushbottom
\maketitle
%
%
\thispagestyle{empty}

\section*{Introduction}

Entropy is an important fundamental property of  many-body systems.
It governs their thermodynamics, heat transfer,
thermoelectric and thermo-magnetic properties. On the other hand, the entropy was always hard to be directly
measured experimentally. It has been revealed very recently that the entropy per particle,
$\partial S/\partial n$, where $n$ is the electron density, can be experimentally studied  \cite{Pudalov2015NatComm}.
To be more precise, the measured quantity is the temperature derivative of the chemical potential, $\partial \mu /\partial T$,
which may be extracted by modulating the temperature of the gated structure with a 2D electron gas
playing the role of one of the plates of a capacitor. Both derivatives are equal as a consequence of the
Maxwell relation
\begin{equation}
\label{entropy-part}
s = \left( \frac{\partial S}{\partial  n} \right)_T = -\left( \frac{\partial \mu }{\partial T} \right)_n.
\end{equation}

In the theoretical paper \cite{Varlamov2016PRB}, quite surprisingly, it has been pointed out that in
a quasi-two-dimensional electron gas (2DEG) with parabolic dispersion the entropy per electron, $s$,
exhibits quantized peaks at resonances between the chemical potential and size quantization levels.
The amplitude of such peaks in the absence of scattering depends only on the
subband quantization number and is independent of material parameters, shape of the confining potential, electron effective mass,
and temperature.

The quantization of entropy per electron was interpreted in \cite{Varlamov2016PRB}
as a signature of the Lifshitz electronic topological transition  \cite{Lifshitz1960JETP},
which in the 2D case is characterised by a discontinuity in the electronic density of states (DOS).
The latter is caused by a change of the topological properties, viz.
connectivity of the electronic Fermi surface \cite{Blanter1994PR}.
Lifshitz transitions widely occur in multi-valley semimetals, doped semiconductor quantum wells,
multi-band superconducting systems such as iron-pnictide compounds \cite{Rodriguez2016JPCM}
and also in 2D Dirac materials, as we discuss below.

In this Report, we analyze theoretically the behavior of
the entropy per particle as a function of the chemical potential in a gapped graphene deposited on a substrate and
other low-buckled Dirac materials, e.g. silicene and germanene.
We show that  the entropy per electron in these systems acquires quantized universal values
at low temperatures if the  chemical potential passes through  the edge of consequent gaps.
It is a universal property of electronic systems characterised by a step-like behaviour of the density of states.
If the chemical potential is resonant to the Dirac point,
we find the discontinuity in $s$ at very low temperature.
At low but finite temperatures this discontinuity transforms into the combination of a very sharp dip at the
negative chemical potential followed by a sharp peak at the positive chemical potential.
These predictions offer a new tool for the characterisation of novel crystalline structures.
In particular, the very characteristic spikes of entropy that must be relatively easy to observe
are indicative of the consequent gaps, in particular due to spin-orbit interaction. We believe that the measurements of
the entropy per particle (e.g. following the technique of Ref. \cite{Pudalov2015NatComm}) may reveal hidden peculiarities
of the band structure of new materials.

\section*{Results}

\subsection*{The link between the discontinuity of the DOS and the quantization of entropy}

To start with, let us consider an electronic system characterised by a DOS function $D (\epsilon)$ that has a discontinuity.
In order to describe Dirac materials specifically, we assume that
the DOS is a symmetric function, $D (\epsilon) =D ( -\epsilon)$, although this assumption is not essential.
We shall assume that the DOS has $2N$ discontinuities at the points $\epsilon = \pm\Delta_i$
and it can be presented in the form
\begin{equation}
\label{DOS-general}
D\left( \epsilon\right) =f(\epsilon)\sum_{i=1}^N \theta \left(\epsilon^{2}-\Delta_i^{2}\right).
\end{equation}
The function
$f(\epsilon)$ is assumed to be a continuous even function of energy $\epsilon$ and it may
account for the renormalizations due to electron-electron interactions in the system.

The case of $N=1$ corresponds to
a gapped graphene with the dispersion law $\epsilon(k) = \pm \sqrt{\hbar ^{2} v_{F}^{2} k^{2} + \Delta ^{2}}$
and $f(\epsilon) = 2 |\epsilon|/(\pi  \hbar ^{2} v_{F}^{2})$, where we have taken into consideration both the valley and spin degeneracy.
Here $ \Delta $ is the gap, $v_{F}$ is the Fermi velocity,
$k$ is the wavevector. The global  sublattice asymmetry gap $2 \Delta \sim 350 \, \mbox {K}$ can be introduced in graphene \cite{Hunt2013Science,Woods2014NatPhys,Chen2014NatCom,Gorbachev2014Science} if it is placed on top of
a hexagonal boron nitride (G/hBN) and the crystallographic axes of graphene and hBN are aligned.

The case of $N=2$ corresponds to
silicene \cite{Kara2012SSR}, germanene \cite{Acun2015JPCM} and other low-buckled Dirac materials \cite{Liu2011PRL,Liu2011PRB}.
The dispersion law in these materials writes $\epsilon_{\eta \sigma} (k) = \pm \sqrt{\hbar ^{2} v_{F}^{2} k^{2} + \Delta_{\eta \sigma} ^{2}}$,
where $\eta$ and $\sigma$ are the valley and spin indices, respectively. Here the valley- and spin-dependent gap,
$\Delta _{\eta \sigma }=\Delta _{z}-\eta \sigma \Delta _{\text{SO}}$, where $\Delta _{\text{SO}}$ is the
material dependent spin-orbit gap caused by a strong intrinsic spin-orbit interaction.
It has a relatively large value, e.g. $\Delta _{\text{SO}} \approx \SI{4.2}{meV}$ in silicene and
$\Delta _{\text{SO}} \approx \SI{11.8}{meV}$ in germanene.
The adjustable gap $\Delta_z = E_z d$, where $2 d$ is the separation between the two sublattices situated in different  planes,
can be tuned by applying an electric field $E_z$.
The function $f(\epsilon) =  |\epsilon|/(\pi  \hbar ^{2} v_{F}^{2})$ is twice smaller than one for graphene, because
the first transition in Eq.~(\ref{DOS-general}) with
$i=1$ corresponds to $\eta = \sigma = \pm$ with $\Delta_1 = | \Delta _{\text{SO}} - \Delta _{z} |$ and
the second one with $i=2$ corresponds to $\eta = -\sigma = \pm$ with $\Delta_2 = |\Delta _{z}+ \Delta _{\text{SO}} |$.

Since the DOS is a symmetric function, instead of the total density of electrons it is convenient to operate with
the difference between the densities of electrons and holes (see the Methods) given by
\begin{equation}
\label{number-general}
 n(T,\mu,\Delta_1,\Delta_2,\ldots,\Delta_N)=
\frac{1}{4} \int _{ -\infty }^{\infty }d \epsilon D (\epsilon) \left [\tanh  \frac{\epsilon +\mu }{2 T} -
\tanh  \frac{\epsilon -\mu }{2 T} \right ],
\end{equation}
where we set $k_B=1$.
Clearly, $n (T ,\mu )$ is an odd function of $\mu $ and $n (T ,\mu  =0) =0$.
The  density $n$ in the Dirac materials may be controlled by an applied gate voltage. In what follows we
consider the dependence of $s$ on the chemical potential.

As it was mentioned above, the entropy per particle is directly related to the
temperature derivative of the chemical potential at the fixed density $n$
(see Eq.~(\ref{entropy-part})). The latter can be obtained using the thermodynamic
identity
\begin{equation}
\label{derivative}
\left( \frac{\partial \mu}{\partial T} \right)_n = - \left( \frac{\partial n}{\partial T} \right)_\mu
\left( \frac{\partial n}{\partial \mu} \right)_T^{-1}.
\end{equation}
If the chemical potential is situated between the discontinuity points, $\Delta_i < |\mu| < \Delta_{i+1}$,
and  $T \to 0$,  one obtains
for the first derivative in Eq.~(\ref{derivative})
(see the Methods)
\begin{equation}
\label{dNdT}
\frac{\partial n(T,\mu)}{\partial T}= D^\prime (|\mu|)\frac{\pi^2 T}{3} \,  {\rm sign}(\mu),
\quad \Delta_i > 0.
\end{equation}
On the other hand,  at the discontinuity points $\mu=\pm\Delta_J$ at $T \to 0$, one finds
\begin{equation}
\label{dNdT-disc}
 \left. \frac{\partial n(T,\mu)}{\partial T} \right|_{\mu=\pm\Delta_J}  =  \pm \left[
D(\Delta_J+0)-D(\Delta_J-0)\right]  \int\limits_{0}^\infty \frac{x \, d x}{\cosh^2x}
 = \pm f(\Delta_J)\ln2.
\end{equation}
One can see that a factor of $\ln 2$ originates from the integration of the derivative of the Fermi distribution
(or $\frac{1}{2} \tanh z $) multiplied by the energy.
If $\mu=\pm\Delta_J$ with $J<N$ and $T \to 0$ for the second derivative in Eq.~(\ref{derivative}), one obtains
(see the Methods)
\begin{equation}
\label{dNdmu}
\left.\frac{\partial n(T,\mu)}{\partial\mu}\right|_{\mu=\pm\Delta_J}   = f(\Delta_J)\sum_{i=1}^N\theta(\Delta^2_J -\Delta^2_i)
 =f(\Delta_J)(J-1/2),
\end{equation}
where the first $J-1$ $\theta$ functions give $J-1$ and the last one gives the $1/2$ contribution.

Thus, we arrive to the conclusion that the entropy per particle in Dirac materials is
\begin{equation}
s(T \to 0,\mu=\pm\Delta_J)=\pm \frac{\ln2}{J-1/2}, \quad J=1,2, \ldots N,
\end{equation}
while for $\Delta_i < |\mu| < \Delta_{i+1}$ it vanishes.
One can see that the behaviour of entropy per particle for the gapped Dirac systems
as a function of chemical potential is analogous to one found in quasi-2DEG with a
parabolic dispersion \cite{Varlamov2016PRB}. This fact allows us to speculate
that such universal spikes are related rather to the topological changes of the Fermi surface
than to specific form of the spectrum.

\subsection*{Gapped Dirac materials}

In the particular case of a gapped graphene the integral (\ref{number-general}) can be done analytically
\cite{Gorbar2002PRB}
\begin{equation}
\label{number-graphene}
n (T ,\mu,\Delta )=  \frac{2 T^{2}}{\pi \hbar ^{2}v_{F}^{2}}\left[ \frac{\Delta }{T}\ln \frac{%
1+\exp \left( \frac{\mu -\Delta }{T}\right) }{1+\exp \left( -\frac{\mu
+\Delta }{T}\right) }
+\mbox{Li}_{2}\left( -e^{-\frac{\mu +\Delta }{T}%
}\right) -\mbox{Li}_{2}\left( -e^{\frac{\mu -\Delta }{T}}\right) \right],
\end{equation}
where $\mbox{Li}$ is the polylogarithm function. The  derivatives $\left( \partial n/\partial T \right)_\mu$
and $\left( \partial n /\partial \mu \right)_T$ are calculated in the Methods, Eqs.~(\ref{derivative-mu}) and (\ref{derivative-T-2nd}).

The density of carriers in silicene can be described with use of the formalism developed above for
graphene by formally representing silicence as a superposition of two gapped graphene layers characterised by different gaps:
$
n  (T ,\mu,\Delta_1, \Delta_2  ) = 1/2 \left[ n (T ,\mu,\Delta_1 ) + n (T ,\mu,\Delta_2 ) \right].
$

\begin{figure}[ht]
\centering
\includegraphics[width=0.99\linewidth]{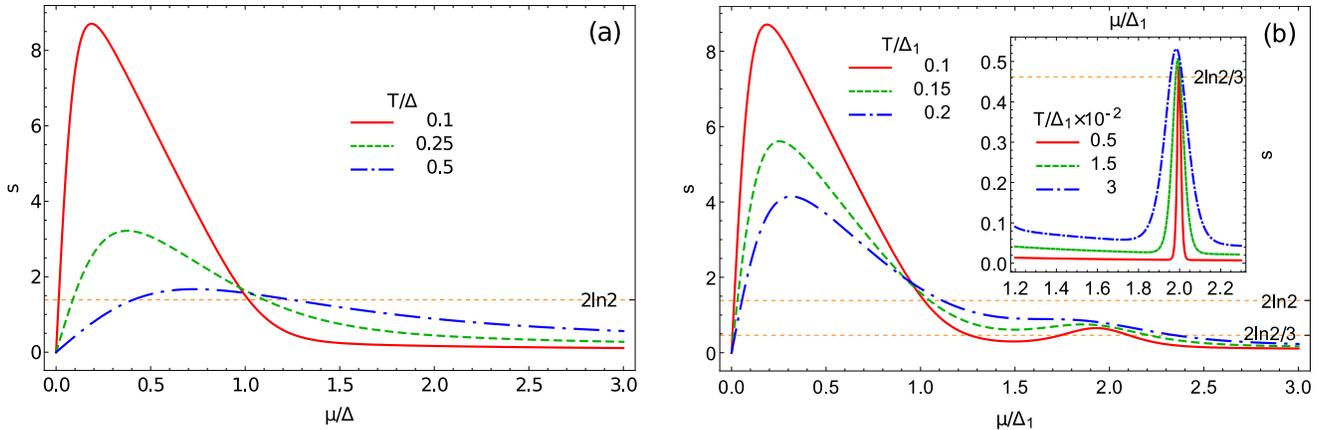}
\caption{{\bf The entropy per electron $s$ vs the chemical potential $\mu >0 $, $s(-\mu) = -s(\mu)$, for three values of temperature.}
Left panel: (a): Gapped graphene. The chemical potential $\mu$ is expressed in the units of $\Delta$; the solid (red) $T/\Delta = 0.1$,
dashed (green) $T/\Delta = 0.25$, dash-dotted (blue) $T/\Delta = 0.5$.
Right panel: (b): Silicene. $\mu$ is in the units of a smaller gap $\Delta_1$, the second  gap
$\Delta_2 = 2 \Delta_1$;  the solid (red) $T/\Delta_1 = 0.1$,
dashed (green) $T/\Delta_1 = 0.15$, dash-dotted (blue) $T/\Delta_1 = 0.2$. The vicinity of
$\mu = \Delta_2$ is shown in the insert:
the solid (red) $T/\Delta_1 = 5 \times 10^{-3}$, dashed (green) $T/\Delta_1 = 1.5 \times 10^{-2}$,
dash-dotted (blue) $T/\Delta_1 = 3 \times 10^{-2}$.  }
\label{fig:1}
\end{figure}

Once the carrier imbalance function, $n(T,\mu,\Delta_1,\Delta_2,\ldots,\Delta_N)$, is found, the entropy per electron
can be calculated using Eqs.~(\ref{entropy-part})
and (\ref{derivative}).
In Fig.~\ref{fig:1} (a) and (b)
we show the dependence $s(\mu)$ for graphene and silicene, respectively,
for three different values of $T$.
Since the entropy per electron is an odd function of $\mu$,
only the region $\mu >0$ is shown.
In the case of silicene we express $\mu$ and $T$
in the units of a smaller gap, $\Delta_1$. The dependence $s(\mu)$ in the vicinity of the second gap,
$\mu = \Delta_2 = 2 \Delta_1$ is shown in the insert of Fig.~\ref{fig:1} (b) to resolve the spike structure
for three temperatures lower than the values on the main plot.

The most prominent feature that we find in Fig.~\ref{fig:1} (a) and (b) is a sharp peak
observed for the chemical potential in the temperature vicinity of the Dirac point,  $|\mu| \sim T$.

If the chemical potential is inside the gap but it is not very close to the Dirac point,
$T \ll |\mu|<\Delta$, and $T \ll \Delta-|\mu|$,
the entropy per particle in a gapped graphene is
\begin{equation}
\label{s-mu<Delta}
s(T, \mu,\Delta  )\simeq\mathrm{sign}(\mu )\left[\frac{\Delta-|\mu|}{T}+1+\frac{T}{\Delta+T}
\right].
\end{equation}
Near the Dirac point, $|\mu| \ll T \ll \Delta$,  one finds
\begin{equation}
\label{mu=0}
s(T, \mu ,\Delta  )\simeq \frac{\mu \Delta}{T^2} \left[ 1 + O (e^{-\Delta/T}) \right].
\end{equation}
If the chemical potential crosses the Dirac point at $T=0$, the transition from hole-like to electron-like
carriers is singular. Eqs.~(\ref{s-mu<Delta}) and (\ref{mu=0}) show how the temperature smears it.
The peak inside the gap is mainly due to the specific dependence of the chemical potential on the electron density.
Indeed, since $s= \partial S(T,\mu)/\partial n =  (\partial S(T,\mu)/\partial \mu)(\partial \mu/ \partial n) $,
the dependence $s(\mu)$ is governed by the sharpest function in the product.
The chemical potential grows rapidly  at the small density $n$ and then quickly reaches the value
$|\mu| \simeq \Delta$, where the derivative
$\partial \mu/ \partial n$ becomes small.
The peaked behavior of $s$ may be considered as a smoking gun for the gap opening in gapped Dirac materials.

Near the Lifshitz transition points: $\mu = \pm \Delta$, we observe that the dependences $s(\mu)$ are monotonic
functions, so that these points are not marked by spikes.
This is typical for any system where DOS has just one discontinuity
\cite{Varlamov2016PRB}.
Nevertheless, the entropy per particle quantization rule for graphene
$s(\mu = \pm \Delta) = \pm 2 \ln2$ is fulfilled.
One can see that in both panels of Fig.~\ref{fig:1}, at low temperatures all  curves cross each other near this point.
The corresponding value $s= 2 \ln 2$ is shown  by the dotted line.
This numerical result can be confirmed analytically.
For $T \ll \Delta$ we obtain
\begin{equation}
\label{2ln2-T}
s(T,\mu=\Delta,\Delta)=2\ln2+\frac{\pi^2-12\ln^22}{3}\frac{T}{\Delta}+
O\left(T^2\right).
\end{equation}

Now we briefly discuss the effect of broadening of the energy levels due to
the scattering from  static defects.
Let us smear the DOS function (\ref{DOS-general})
by convoluting it with the Lorentzian, $\gamma/[\pi(\omega^2+\gamma^2)]$,
where $\gamma$ is the scattering rate.
In the regime $\gamma\ll T\ll\Delta$ one finds
\begin{equation}
\label{2ln2-dirty}
s(T, \mu = \Delta, \Delta)=2\ln2 \left[1-\frac{\gamma}{T}\left(\frac{1}{\pi\ln2}
+\frac{T}{\Delta}\right)\right].
\end{equation}
Eq.~(\ref{2ln2-dirty}) shows that the universality of the low temperature entropy per particle
is broken by the disorder if the mean free path becomes comparable with the thermal diffusion length.

The case $\Delta=0$ deserves a special attention. In this limit, Eq.~(\ref{number-graphene})
acquires a simple form (see the Methods, Eqs.~(\ref{number-graphene-0}) and (\ref{derivative-Delta=0})).
For the entropy per particle one finds
\begin{equation}
\label{s(T,0,mu)}
s(T,\mu,0)=\left\{%
\begin{array}{cc}
\frac{\mu}{T}\left(1-\frac{\mu^2}{T^2}\frac{1}{6\ln2}\right), \quad |\mu| \ll T,
&  \\
\frac{\pi^2}{3}\frac{T}{\mu},\quad T\ll |\mu| . &
\end{array}%
\right.
\end{equation}
It is important to note that the second line of Eq.~(\ref{s(T,0,mu)})
if multiplied by the factor $k_B/e$ yields the Seebeck coefficient for a free electron gas \cite{Abrikosov.book}.
Moreover, the general expression for $s = -\partial \mu/\partial T$, Eq.~(\ref{derivative-Delta=0})
reproduces the thermal power $S$ that can be extracted  from the results based on the Kubo formalism \cite{Sharapov2012PRB}
that validates  the thermodynamic approach of \cite{Varlamov2013EPL}.

The presence of the second gap in silicene and similar materials, $\Delta_2 > \Delta_1$, results in the appearance of the
peak in $s(\mu) \approx \pm  2 \ln2/3$ near the point $\mu = \pm \Delta_2$, as seen in Fig.~\ref{fig:1}(b).
The corresponding value $s= 2 \ln2/3$ is shown by the dotted line.
This peak can be considered as a signature of the second Lifshitz transition which occurs  if $\mu$
crosses $\Delta_2$. Indeed, as it was shown for the quasi-2DEG in \cite{Varlamov2016PRB}
the peak structure in $s(\mu)$ develops only if the number of discontinuities in the DOS, $N \geq 2$.
Thus, these perspective Dirac materials, where the spin orbit interaction plays a very important role
allow the simplest realization of the $N = 2$ case with two discontinuities on  both electron
and hole sides of the total DOS.

\begin{figure}[ht]
\centering
\includegraphics[width=0.99\linewidth]{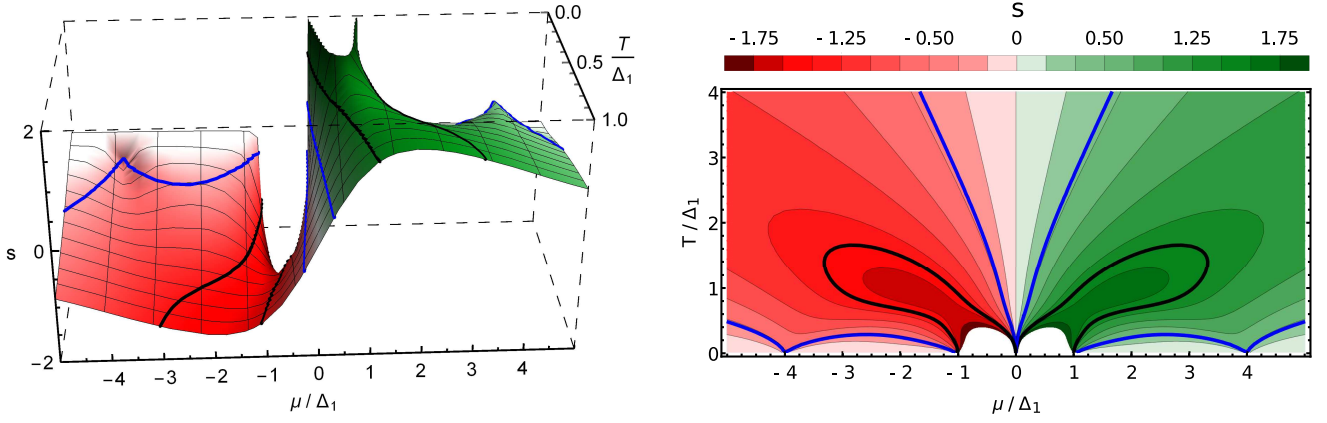}
\caption{{\bf The entropy per electron $s$
as functions of the chemical potential $\mu$ and temperature $T$ in the units of $\Delta_1$.} The gap $\Delta_ 2 = 4\Delta_1 $.
Left panel:  3D plot.  Right panel:  Contour plot. }
\label{fig:2}
\end{figure}

Fig.~\ref{fig:2} shows the 3D and density plots of $s$ as a function of
$\mu/\Delta_1$ and $T/\Delta_1$.
To be specific, we assumed that $\Delta_1$ is the smallest of the gaps and chose $\Delta_2 = 4 \Delta_1$.
The black and blue lines correspond to the contours of constant values
$s = \pm 2 \ln 2$ and $s = \pm 2 \ln 2 /3$, respectively.
The range of $s$ in the 3D plot is restricted by $-2 \leq s \leq 2$, so that only the peaks at $\mu =\pm \Delta_2$ can be
seen.

A more careful examination of Fig.~\ref{fig:1}~(b) shows that the peak occurring near  $\mu = \Delta_2$
is somewhat shifted to smaller than $\Delta_2$ values of $\mu$. Looking at Fig.~\ref{fig:2}~(b) and its insert
one can trace  how the position of this peak moves towards the point $(\mu = \Delta_2, T =0)$
as the temperature decreases. In Fig.~\ref{fig:2}~(a)  the increase of its height can be seen.
Close to this point ($T \ll \Delta_2$) we obtain analytically
\begin{equation}
\label{2ln2-T-2nd}
s(T,\mu=\pm\Delta_2)=\pm\left[\frac{2\ln2}{3}+\frac{\pi^2-4\ln^22}{9}\frac{T}{\Delta_2}\right].
\end{equation}
In what concerns the behaviour the silicene's entropy per particle close to the smallest gap, $\Delta_1$,
it is described by  Eq.~(\ref{2ln2-T}) with  $\Delta$ replaced by $\Delta_1$.

\begin{figure}[ht]
\centering
\includegraphics[width=0.99\linewidth]{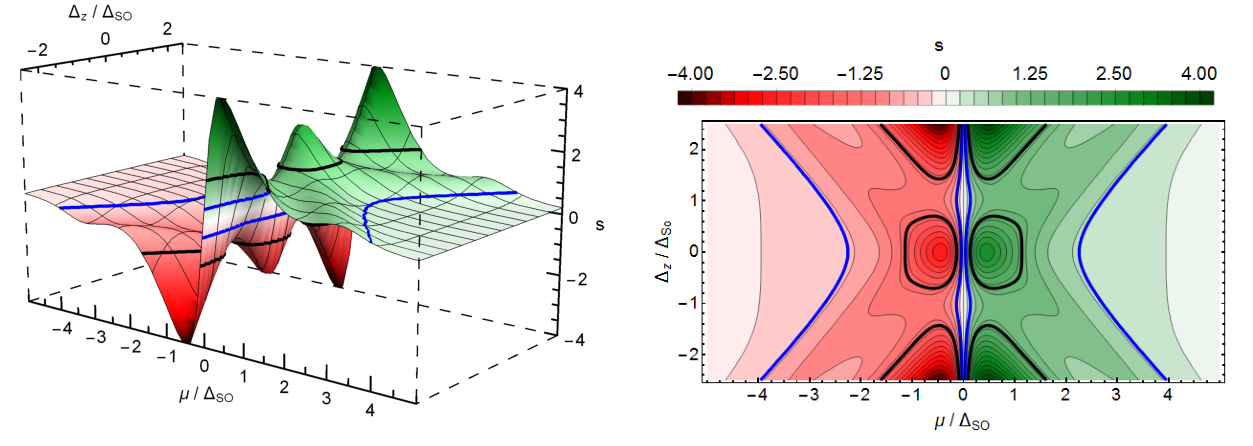}
\caption{{\bf The entropy per electron $s$
as functions of the chemical potential $\mu$ and $\Delta_z$  in the units of $\Delta_{\mathrm{SO}}$. The temperature $T =0.3 \Delta_{\mathrm{SO}}$.}
Left panel:  3D plot.  Right panel:  Contour plot. }
\label{fig:3}
\end{figure}

Recent successes in fabrication of silicene field-effect transistors \cite{Tao2015NatNano}
offers the opportunity of a direct measurement of the entropy per particle in silicene.
In the prospective experiment, a double gate structure would be needed that enables one to tune $\mu$ and $\Delta_z$
independently. Such a situation is modelled in Fig.~\ref{fig:3}, where we  show the 3D and density  plots of $s$ as a function of
$\mu/\Delta_{\mathrm{SO}}$ and $\Delta_z/\Delta_{\mathrm{SO}}$.
As in Fig.~\ref{fig:2}, the black and blue lines correspond to the contours of constant values
$s = \pm 2 \ln 2$ and $s = \pm 2 \ln 2 /3$, respectively.
The points $\Delta_z = \pm \Delta_{\mathrm{SO}}$ correspond to the case where
$\Delta_1 =0$ and $\Delta_2 = 2 \Delta_{\mathrm{SO}}$ or $\Delta_1 = 2 \Delta_{\mathrm{SO}}$ and $\Delta_2 =0$, so that
the system experiences a transition from two to one gap spectrum.
For $|\Delta_z|  < \Delta_{\mathrm{SO}}$ the system is a topological insulator
and for $|\Delta_z|  > \Delta_{\mathrm{SO}}$ it is a band insulator.

\section*{Discussion}

We presented original analytical expressions for the entropy per particle in a wide energy range for various Dirac materials.
Basing on them we have predicted the characteristic spikes of the entropy per particle at the
Lifshitz topological transition points in several 2D Dirac systems.
The magnitude of spikes is quantized at low temperatures and is independent of material parameters.
The quantized spikes are expected to occur in silicene and germanene. They can also be found in the
gapped graphene in the presence of Zeeman splitting and in quasi two-dimensional Dirac and Weyl materials.
Note that the same quantization of entropy and spikes occur in a 2DEG  in
the presence of Zeeman splitting  \cite{Pudalov2015NatComm}, see the Methods.

Our results are based on the assumption that the function $f(\epsilon)$ in the
DOS (\ref{DOS-general}) is continuous. Although this assumption is quite
general, it is not fulfilled, for example, in  a bilayer graphene.
The overall behavior of the entropy per electron $\partial S/\partial n$ as a function of the electronic chemical potential
may be used as a tool for characterization of the electronic dispersion in novel crystal structures.
The crucial point is that $\partial S/\partial n$ is related to the temperature derivative
$\partial \mu/\partial T$ via the thermodynamic  Maxwell relation
(\ref{entropy-part}). The last value, as was mentioned in Introduction, can be directly measured using the experimental
approach  developed in \cite{Pudalov2015NatComm}. It appears that this technique has a three orders of
magnitude higher resolution than the other methods and thus it can be very helpful in probing interaction effects
in 2D electron systems.
The measurements of the entropy per particle can also be used to study
the effect of interactions on the DOS in graphene, because
the renormalization of the Fermi velocity due to electron-electron interactions \cite{Elias2011NatPhys}
modifies the function $s (n)$.

\section*{Methods}

\subsection*{Relationship between the carrier density and carrier imbalance}

At thermal equilibrium, the total density of electrons
in a nonrelativistic system can be expressed as
\begin{equation}
n_{\mathrm{tot}}(T,\mu)  =\int _{-\infty}^{\infty }d \epsilon D (\epsilon ) f_{F D} \left(\frac{\epsilon -\mu }{T} \right),
\end{equation}
where $f_{F D} ( \epsilon ) = 1/[\exp(\epsilon/ T)+1 ]$ is the Fermi-Dirac distribution function and we set $k_B=1$.
In a relativistic theory, for example, in  QED the number of electrons or positrons is not conserved, while
a conserving number operator is needed to build the statistical density matrix \cite{Kapusta.book}.
In QED, the conserved quantity if the difference of the numbers of positively and negatively charged particles:
electrons and positrons.

In the Dirac materials the ``relativistic'' nature of carriers is encoded in the
symmetric DOS function, $D(\epsilon) = D(-\epsilon)$. Accordingly,
it is convenient to operate with the difference between the densities of electrons and holes instead of the total
density of electrons \cite{Gusynin2004PRB,Sharapov2015JPA}. The difference
is given by
\begin{equation}
\label{number-rel}
n(T,\mu)  =  \int_{-\infty}^{\infty } d \epsilon D(\epsilon) [ f_{F D}(\epsilon - \mu) \theta(\epsilon)
 -[1- f_{F D}(\epsilon - \mu) ] \theta(-\epsilon)]
 = -\frac{1}{2} \int_{-\infty}^{\infty } d \epsilon D(\epsilon) \tanh \frac{\epsilon-\mu}{2T}.
\end{equation}
The last equation can be rewritten in the form of Eq.~(3).
One can verify that the carrier imbalance $n(T,\mu)$ and the total carrier density $n_{\mathrm{tot}}(T,\mu) $
are related by the expression
$n(T,\mu)  = n_{\mathrm{tot}}(T,\mu) - n_{\mathrm{hf}}$, where $n_{\mathrm{hf}}$ is the density of particles
for a half-filled band (in the lower Dirac cone)
$
n_{\mathrm{hf}} =  \int_{-\infty}^{\infty } d \epsilon D(\epsilon) \theta(-\epsilon).
$
Consequently, there is no difference whether the entropy per particle in Eq.~(\ref{entropy-part}) is defined
via the total carrier density $n_{\mathrm{tot}}$ or the carrier imbalance $n$.

\subsection*{General expressions for $ \partial n /\partial T$ and $\partial n /\partial \mu$}

The first temperature derivative in Eq.~(\ref{derivative}) depends on whether the chemical potential
$\mu$ hits the discontinuity of the DOS $D(\epsilon)$ given by Eq.~(\ref{DOS-general}).
Differentiating Eq.~(\ref{number-general}) over the temperature one obtains
\begin{equation}
 \frac{\partial n(T,\mu)}{\partial T} =  \frac{{\rm sign}(\mu)}{4T}\int_{-\infty}^\infty d\epsilon D(\epsilon)
 \left[
\frac{\epsilon-|\mu|}{2T}\frac{1}{\cosh^2\frac{\epsilon-|\mu|}{2T}}-\frac{\epsilon+|\mu|}{2T}\frac{1}{\cosh^2
\frac{\epsilon+|\mu|}{2T}}\right].
\end{equation}
Changing the variable $\epsilon=2Tx\pm|\mu|$ in two terms and changing the limits of integration, one obtains
\begin{equation}
 \frac{\partial n(T,\mu)}{\partial T} =
 {\rm sign}(\mu)\int\limits_{0}^\infty dx\left[D(|\mu|+2Tx)-D(|\mu|-2Tx)\right]
\frac{x}{\cosh^2x}.
\end{equation}
If the DOS $D(\epsilon)$  has a continuous derivative at the point $\epsilon = |\mu|$, where $\Delta_i < |\mu| < \Delta_{i+1}$,
one can expand $D(|\mu|+2Tx)-D(|\mu|-2Tx)\simeq 4TxD^\prime(|\mu|)$. Then integrating over $x$ we arrive at Eq.~(\ref{dNdT})
\begin{equation}
\frac{\partial n(T,\mu)}{\partial T}  \simeq4T{\rm sign}(\mu)D^\prime(|\mu|)\int\limits_{0}^\infty \frac{x^2 \, d x}{\cosh^2x}
={\rm sign}(\mu)D^\prime(|\mu|)\frac{\pi^2}{3}T.
\end{equation}
On the other hand,  at the discontinuity points $\mu=\pm\Delta_J$ at $T \to 0$, we arrive at Eq.~(\ref{dNdT-disc}).

The second derivative in Eq.~(\ref{derivative}) in the zero temperature limit is just the DOS.
Indeed, we have
\begin{equation}
\label{mu-derivative}
\frac{\partial n(T,\mu)}{\partial\mu} =\frac{1}{8T}\int_{-\infty}^\infty d\epsilon D(\epsilon)\left[
\frac{1}{\cosh^2\frac{\epsilon+\mu}{2T}}+\frac{1}{\cosh^2\frac{\epsilon-\mu}{2T}}\right]
 =  D(\mu),\qquad T\to0.
\end{equation}
This is because
$(1/4T) \cosh^{-2} (x/2T) \rightarrow\delta(x)$ for $x\to0$. Substituting the DOS given by Eq.~(\ref{DOS-general}) to
Eq.~(\ref{mu-derivative}) we arrive at Eq.~(\ref{dNdmu}).

\subsection*{Explicit expressions for the derivatives $ \partial n /\partial T$ and $\partial n /\partial \mu$ for the Dirac materials}

The carrier imbalance for a gapped graphene is given by Eq.~(\ref{number-graphene}). The corresponding derivatives
are
\begin{equation}
\label{derivative-mu}
 \left( \frac{\partial n}{\partial \mu }\right) _{T} =  \frac{2}{\pi \hbar
^{2}v_{F}^{2}}\left[ \frac{\Delta }{2}\left( \tanh \frac{\mu -\Delta }{2T}%
-\tanh \frac{\mu +\Delta }{2T}\right)
+ T\left( \ln \left( 2\cosh \frac{\mu -\Delta }{2T}\right) +\ln
\left( 2\cosh \frac{\mu +\Delta }{2T}\right) \right) \right]
\end{equation}
and
\begin{equation}
\label{derivative-T-2nd}
\begin{split}
 \left( \frac{\partial n}{\partial T} \right)_\mu = & \frac{2}{ \pi \hbar^2 v_F^2}
\left[ 2\Delta \ln \frac{1+\exp \left( \frac{\mu -\Delta }{T}%
\right) }{1+\exp \left( -\frac{\mu +\Delta }{T}\right) }
 +2T\mbox{Li}_{2} \left( -e^{-\frac{\mu +\Delta }{T}}\right)
-2T\mbox{Li}_{2}\left( -e^{\frac{\mu -\Delta }{T}}\right) \right. \\
& \left. -\mu \ln \left( 2\cosh \frac{\mu -\Delta }{2T}\right) -\mu \ln \left( 2\cosh \frac{%
\mu +\Delta }{2T}\right)
+ \frac{\Delta }{T}\frac{\mu \sinh (\Delta /T)+\Delta \sinh \mu /T}{%
\cosh \Delta /T+\cosh \mu /T}\right] .
\end{split}
\end{equation}
Eqs.~(\ref{s-mu<Delta}) -- (\ref{2ln2-T}) and (\ref{2ln2-T-2nd}) are obtained using the low-temperature expansions of the derivatives,
Eqs.~(\ref{derivative-mu}) and (\ref{derivative-T-2nd}).

\subsection*{Dirac materials with $\Delta=0$}

If $\Delta =0$ Eq.~(\ref{number-graphene}) reduces to
\begin{equation}
\label{number-graphene-0}
n(T,\mu) = \frac{2T^2}{\pi \hbar^2 v_F^2}
\left[ \mbox{Li}_2 \left(-e^{- \frac{\mu}{T}} \right) - \mbox{Li}_2 \left(-e^{ \frac{\mu}{T}} \right) \right].
\end{equation}
Using Eq.~(\ref{derivative}) we obtain the general expression
\begin{equation}
\label{derivative-Delta=0}
\left( \frac{\partial \mu}{\partial T} \right)_n  = \frac{\mu}{T}-  \frac{1}{\ln \left (2 \cosh \frac{\mu}{2T} \right)}
\left[ \mbox{Li}_2 \left(-e^{- \frac{\mu}{T}} \right) - \mbox{Li}_2 \left(-e^{ \frac{\mu}{T}}\right) \right].
\end{equation}

\subsection*{Quantization of entropy in the presence of Zeeman splitting}

In the 2DEG in the presence of Zeeman splitting considered in the Supplementary material of \cite{Pudalov2015NatComm}
the carrier density reads
\begin{equation}
\label{number:eq}
n(\mu,T)=\frac{m}{4\pi}T\left[\ln\left(1+e^{(\mu+Z)/T}\right)+\ln\left(1+e^{(\mu-Z)/T}\right)\right].
\end{equation}
Here $Z$ is the Zeeman splitting energy and $m$ is the carrier mass.
One can show that the entropy per particle in this case also obeys
the  quantization rule
\begin{equation}
\label{s-minus-Z}
\left. \frac{\partial S}{\partial n} \right|_{\mu=-Z}=2\ln2,
\quad \left. \frac{\partial S}{\partial n} \right|_{\mu=Z}=\frac{2\ln2}{3}, \quad T\to 0.
\end{equation}

\section*{Acknowledgements }

We acknowledge the support of EC for the RISE Project CoExAN GA644076.
A.V.K acknowledges support from the EPSRC established career fellowship.
V.P.G. and S.G.Sh. acknowledge a partial support from
the Program of Fundamental Research of the Physics and Astronomy
Division of the NAS of Ukraine  No. 0117U00240.

\section*{Author contributions statement}

A.V.K., S.G.Sh., A.A.V. and V.P.G. conceived the work. S.G.Sh., A.A.V. and V.P.G. performed calculations. V.Yu.T.
has done all numerical computations and prepared the figures.
All authors contributed to writing the manuscript.

\section*{Additional information}

\textbf{Competing financial interests:} The authors declare no competing financial interests.

\end{document}